\documentclass[12pt,draftcls,onecolumn]{IEEEtran}
\usepackage{hyperref}
\usepackage{amsmath,amssymb}
\usepackage{latexsym}
\usepackage{CJK}
\usepackage{cite}   
\usepackage{caption} 

\usepackage{verbatim}  
\usepackage{graphicx}  
\usepackage{bm}  
\usepackage{mathrsfs} 
\usepackage{algorithm} 
\usepackage{algorithmic} 
\usepackage{booktabs}
\usepackage{textcomp}  
\usepackage{multirow}  
\usepackage{lettrine}   
\usepackage{graphicx}  
\usepackage{color}  
\usepackage{amsmath}
\usepackage{amssymb}

\DeclareMathOperator*{\argmin}{argmin}

\usepackage{algorithm}
\usepackage{algorithmic}

\usepackage{amsthm} 

\theoremstyle{plain}

\newtheorem{lem}{Lemma}

\theoremstyle{definition}

\theoremstyle{remark}
\newtheorem*{rem}{Remark}

\newcommand{\myvec}[1]%
   {\stackrel{\raisebox{-2pt}[0pt][0pt]{\small$\rightharpoonup$}}{#1}}

\begin{document}
%
\title{Massive MIMO Beam-forming for High Speed Train Communication: Directivity vs Beamwidth}


\author{Xuhong Chen, Jiaxun Lu, Pingyi~Fan~\IEEEmembership{Senior Member,~IEEE,}\\
        State Key Laboratory on Microwave and Digital Communications, \\
        Tsinghua National Laboratory for Information Science and Technology, \\
        Department of Electronic Engineering, Tsinghua University, Beijing, China.\\
\thanks{Corresponding Author is Pingyi Fan.}
E-mail: chenxh13@mails.tsinghua.edu.cn, lujx14@mails.tsinghua.edu.cn, \\
        fpy@mail.tsinghua.edu.cn}

\maketitle

\begin{abstract}
High-mobility adaption and massive Multiple-input Multiple-output (MIMO) application are two primary evolving objectives for the next generation high speed train communication system. In this paper, we consider how to design a location-aware beam-forming for the massive MIMO system. We first analyze the tradeoff between beam directivity and beamwidth, based on which we present the sensitivity analysis of positioning accuracy. Then, we derive the maximum beam directivity and corresponding beamwidth under the restriction of diverse positioning accuracies to guarantee a high efficient transmission. Finally, we present a low-complexity beam-forming design with positioning robustness utilizing location information, which requires neither eigen-decomposing (ED) the uplink channel covariance matrix (CCM) nor ED the downlink CCM (DCCM). Numerical simulation indicates that a massive MIMO system with less than a certain positioning error can guarantee a required performance with satisfying transmission efficiency in the high-mobility scenario.
\end{abstract}


\IEEEpeerreviewmaketitle

\section{Introduction}
\lettrine[lines=2]{H}igh mobility adaption is a crucial target for the next generation wireless communication system \cite{mobility, mobility1}, where in diverse high mobility scenarios such as high speed train (HST) scenario, the transmission demands is ever-growing according to the real-scenario estimation in \cite{demand} (the estimated demands could be as high as 65Mbps with bandwidth 10MHz for a train with 16 carriages and 1000 seats). In the literature, certain different designs aiming to improve the user quality of service (QoS) in high mobility scenario have been proposed \cite{work, work1}. However, low-complexity beam-forming design for high mobility scenarios are still under-developed.

Massive Multiple-input Multiple-output (MIMO) is deemed as a prominent technology in future 5G system to improve the spectrum efficiency \cite{5G, 5G1} through exploiting the multiplexing and diversity gain, where the appropriately-designed beam-forming is utilized to diminish interference \cite{beamform}. Previous works in \cite{work2, work3, work4} employing diverse beam-forming schemes in high mobility scenario demonstrated a significant performance improvement by utilizing the directional radiation. However, neither of them considered how to reduce the implementation complexity of beam-forming for massive MIMO system in HST scenario, especially when channel estimation is required. That is, the online computational complexity and challenges in the process of channel covariance matrix (CCM) acquisition and eigen-decomposing (ED) CCM when adopting the massive MIMO beam-forming.

\begin{figure}[!b]
\centering
\includegraphics[width=0.45\textwidth]{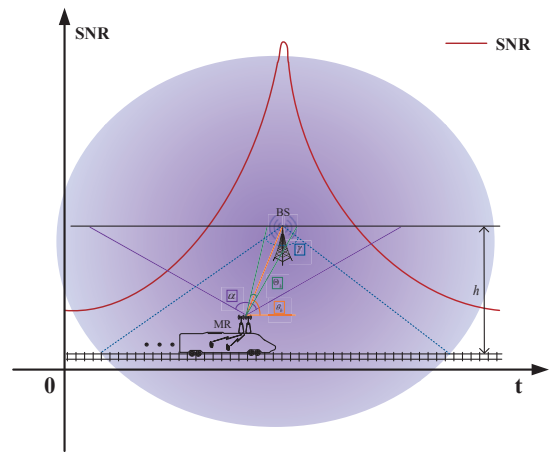}
\caption{The HST scenario and beam coverage.} \label{fig:hsr}
\captionsetup{belowskip=-10pt}
\end{figure}
In the high mobility scenario, the special scenario characters distinguish this scenario from conventional low mobility scenario. In this paper, we take the HST scenario as an example, which is as illustrated in Fig. \ref{fig:hsr}. As the train quickly traversing the coverage area of one base station (BS), the received signal-to-noise ratio (SNR) at the mobile relay (MR) will fluctuate dramatically due to the large variation of path loss, which makes the received SNR in the edge area of the BS coverage quite low, bringing challenges for conventional adaptive beam-forming \cite{abf} or orthogonal switched beam-forming \cite{obf} in channel detections. The estimated maximum Doppler shift with carrier frequency at 2.35 GHz will be 945 Hz when the train velocity is 486 km/h \cite{demand}, which implies that the channel coherence time is less than 1 ms. Consequently, it is hardly to track the channel in this scenario. Moreover, the complex channel environment due to traversing diverse terrains (typical scenarios like viaduct, mountain, etc.) makes it difficult to estimate the channel with low cost since the wireless channel appears fast time-varying and double-selective fading in spatial-temporal domains. On the other hand, even if the channel state information (CSI) is acquired, conventional beam-forming design methods for massive MIMO system in low mobility scenario can not be directly applied here since it will lead to undesirable performance. Therefore, beam-forming scheme for massive MIMO system in this scenario requires to be redesigned with less complicated method.

Although high mobility causes new challenges in the application of beam-forming scheme, if we take advantage of high mobility in a different perspective, the drawback can be transformed into valuable side information, namely, the location of the vehicle can be trackable and predictable because the moving trend will not change in a short time due to the safety guarantee for high speed vehicles. Actually for HST scenario, the train can only move along the pre-constructed rail, which means no spatial-random burst communication requests will occur and it will narrow down the coverage scope of the beam-forming scheme.

In this paper, we introduce a simple low-complexity beam-forming scheme for massive MIMO system in HST scenario by exploiting the vehicle location information. Because the conduction of this beam-forming scheme required neither ED uplink CCM (UCCM) or downlink CCM (DCCM) and therefore, the aforementioned challenges in channel detections and large online computational complexity can be alleviated, that is, the developed scheme can work well for imperfect CSI. Since the train location information plays a crucial role in the beam-forming scheme, we first analyze the tradeoff between beamwidth and directivity in high mobility scenario and find that it is independent of antenna spacing and total beam number. Then, we present the sensitivity analysis of positioning accuracy against beam directivity and formulate it as an optimization problem. Through transforming the optimization problem, an optimal searching-based algorithm is proposed to maximize the directivity, while guaranteeing an efficient transmission at the same time when the positioning error is constrained.

The contributions of this paper can be concluded as
 \begin{itemize}
\item \emph{\textbf{A tradeoff analysis between beam directivity and beamwidth for the beam-forming scheme is presented.}}
\item \emph{\textbf{A location-sensitivity analysis and an optimal beam-forming solution to maximize the directivity for a given error range or error distribution of the location information are provided.}}
\item \emph{\textbf{A location-aware low-complexity beam-forming scheme for a high-mobility massive MIMO system is given.}}
\end{itemize}

The rest of this paper is organized as follows. Section II introduces the system model and the transceiver structure. Section III presents the directivity-beamwidth tradeoff analysis and the optimization solution to maximize the directivity under diverse positioning error constrains. Section IV presents the application process of designed low-complexity beam-forming scheme for HST scenario. Section V shows the numerical simulation results and the corresponding analyses. Finally, we conclude it in Section VI.

\section{System Model}

\begin{figure}[!b]
\centering
\includegraphics[width=0.5\textwidth]{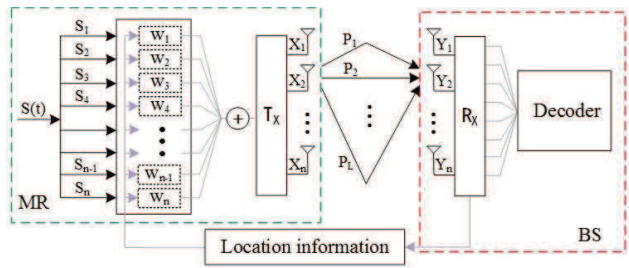}
\caption{The transceiver structure for HST massive MIMO systems.} \label{fig:transceiver}
\captionsetup{belowskip=-10pt}
\end{figure}

Let us consider the beam-forming transceiver structure between one carriage and the BS, depicted in Fig. \ref{fig:transceiver}, where we omit most parts of baseband transformations and only focus on the beam-forming part. The user transmission inside the train will be forwarded by the MR mounted on the top of the train to avoid large penetration loss and the MR is equipped with $M$-element uniform linear single-array antenna due to the \emph{3dB} gain over double-array structure \cite{gain}. To the link between BS and the MR, the Doppler effect in HST scenario will not be considered here since it can be accurately estimated and removed from the signal transmission part as shown in \cite{work1}. Let the antenna spacing and carrier wavelength be $d$ and $\lambda$, respectively. For a multi-element uniform linear array, where $d \gg \lambda$, the half-power beamwidth $\Theta_h$ can be equivalently expressed as \cite{antenna}

\begin{equation}\label{F1}
\Theta_h = \frac{C \lambda}{\pi d N},
\end{equation}
where $C=2.782$ represents a constant parameter in antenna design and $N$ is the total generated beam number.

In HST scenario, the viaduct scenario occupies 80 percent of the entire route \cite{demand}, which makes the line-of-sight (LOS) signal dominant and the angular spread around the MR relatively tight. In addition, the instantaneous location information $\theta_b$ (shown in Fig. \ref{fig:hsr}) of BS can be acquired by Global Positioning Systems, accelerometer and monitoring sensors along the railway or, can be estimated according to the entrance time as \cite{work4}. Thus, a location-aware beam-forming can be carried out by exploiting $\theta_b$ to reduce the beam-forming complexity.

As shown in Fig. \ref{fig:hsr}, the total beamwidth, which is a constant for a deployed BS, can be expressed by
\begin{equation*}
 \alpha=N \Theta_h = \frac{C \lambda}{\pi d}.
\end{equation*}

To enhance the coverage of the MR on the train, $\alpha > \gamma$ is essential, where $\gamma$ is the coverage angle of BS.

\section{Directivity-Beamwidth Tradeoff and Efficient Transmission}\label{Sec:BeamAndConnection}

In practice, positioning error may occur due to some reasons, which may degrade the performance of the location-aware low-complexity beam-forming scheme and trigger low transmission efficiency. In this section, we first analyze the tradeoff between beam directivity and beamwidth and then, derive the efficient beam-forming probability under diverse positioning error constrains, where the effective beam-forming is defined as the BS is in the coverage of the selected beam. Finally, under the precondition of efficient transmission, we maximize the beam directivity with error-constrained location information.

The corresponding directivity of each beam in massive MIMO system, i.e. $N\pi d/\lambda$ is sufficiently large, can be expressed as

\begin{equation}\label{F2}
D = \frac{TdN}{\lambda},
\end{equation}
where $T$ depends on the specific type of linear array (i.e. $T=2$ for broadside array and $T=4$ for ordinary end-fire array).

Thus, the relations between beam directivity and beamwidth can be expressed in the following lemma.

\begin{lem}The tradeoff between beam directivity and beamwidth is

\begin{equation}\label{F3}
D = \frac{TC}{\pi \Theta_h}.
\end{equation}

\end{lem}

\begin{rem}
The tradeoff between beam directivity and beamwidth is independent of the antenna spacing $d$ and the beam number $N$. That is, the variation of $d$ and $N$ only affects the value of beam directivity and beamwidth, but has no influence on the ratio of the beam directivity and beamwidth.
\end{rem}
Observing Eq. (\ref{F1}) -(\ref{F3}), one can find that
\begin{rem}
$d$ and $N$ are dual with respect to $\Theta_h$ and $D$. That is, the variation of $d$ $\rightarrow$ $d^{'}$ is equivalent to $N$ $\rightarrow$ $N^{'}$, where
$$
\frac{d}{d^{'}} = \frac{N^{'}}{N}.
$$
\end{rem}

As shown in Fig. \ref{fig:hsr}, let the BS's location and the vertical distance between BS and rail be $\theta_b$ and $h$, respectively. Thus, the BS is supposed to be in the $i$-th beam, where

\begin{equation}\label{equ:index}
\begin{split}
i &= \biggl\lfloor\frac{2\theta_b + \alpha - \pi}{2\Theta_h}\biggr\rfloor\\
& = \biggl\lfloor\frac{2\theta_b - \pi}{2\Theta_h}\biggr\rfloor + \frac{N}{2}.
\end{split}
\end{equation}

\begin{figure}[b]
\centering
\includegraphics[width=0.43\textwidth]{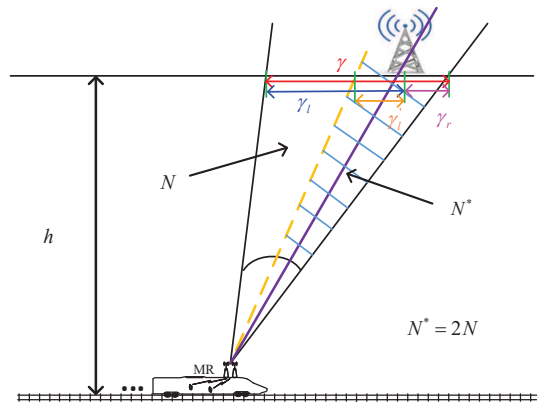}
\caption{The beam devision process.} \label{Fig:BeamDevide}
\end{figure}

Then, as illustrated in Fig .\ref{Fig:BeamDevide}, the distances between left(right) bounds (denoted as $\gamma_l$($\gamma_r$)) of $i$-th beam and BS can be expressed by

\begin{equation}\label{equ:gamma_l}
\gamma_l = \frac{[\pi+2\chi\Theta_h-2\theta_b]h}{2\sin \theta_b}
\end{equation}
and

\begin{equation}\label{equ:gamma_r}
\gamma_r = \frac{[2 \theta_b-\pi-2(\chi-1)\Theta_h]h}{2 \sin \theta_b},
\end{equation}
respectively, where $\chi = \biggl\lfloor\frac{2\theta_b - \pi}{2\Theta_h}\biggr\rfloor$.

Therefore, the corresponding coverage length on the BS side of the $i$-th beam is given by

\begin{equation}
\gamma = \gamma_l + \gamma_r \approx \frac{h}{\sin \theta_b} \Theta_h = \frac{C \lambda h}{\pi d N \sin \theta_b}.
\end{equation}

When there exists positioning error $\Delta x$, the effective beam-forming rate may be decreased. In fact, $\Delta x$ may be caused by GPS estimation errors, quantization errors, random scattering of electronic waves, etc. For simplicity, it is assumed that $\Delta x$ is Gaussian distributed in the sequel, i.e. $\Delta x \sim \emph{N}(0,\sigma^2)$. We shall utilize it in the following analysis and optimization.

With the widely used Q function in signal detection, i.e. $Q(x) = \int_{x}^{+\infty} \frac{1}{\sqrt{2 \pi}} exp(-\frac{t^2}{2})dt$, the effective beam-forming probability $P_i$ can be expressed by

\begin{equation}
P_i = 1 - \frac{Q(\frac{\gamma_l}{\sigma}) + Q(\frac{\gamma_r}{\sigma})}{2}.
\end{equation}

The effective beam-forming probability threshold is denoted as $P_{th}$. Thus, we try to improve the system gain of beamforming and formulate the following optimization problem

\begin{align*}\label{equ:formulation}
\tag{9}&\max \,\, D\\
\tag{10}&s.t.\quad  P_i \geq P_{th}.
\end{align*}

In (9), the object is to maximize the directivity (beam gain) and (10) represents the efficient transmission constraint condition. To solve the maximization problem above, we first analyze the monotonicity of $P_i$. As shown in Fig. \ref{Fig:BeamDevide}, if increasing the beam number from $N$ to $N^{'} = 2N$, and keeping $\gamma_r$ unchanged, then $\gamma_l$ has been shorten to $\gamma_l^{'} = \gamma_l - \frac{\gamma}{2}$ and $P_i^{'} = P_i - \frac{Q(\gamma_l^{'}) - Q(\gamma_l)}{2}$. That indicates the $P_i$ monotonous decreases with respect to $N$. Therefore, the solving of (9) can be turned into a searching problem as follows

\begin{equation}\label{equ:minError}
N^*=\mathop{\argmin}_N (P_i - P_{th}) \tag{11}.
\end{equation}

Note that the directivity $D$ increases in proportion to $N$.

In literature, there are many techniques to solve (11). Considering the fact that number of available $N$ is finite in a system, the searching process can be summarized as follows.

\begin{algorithm}\label{alg:search}
\caption{The searching method of $N^*$}
\begin{algorithmic}
\STATE {set $N^*=1$}
\REPEAT
\STATE 1. generate i, $\gamma_l$ and $\gamma_r$ with \eqref{equ:index}, \eqref{equ:gamma_l} and \eqref{equ:gamma_r}.
\STATE 2. $1 - \frac{Q(\gamma_l) + Q(\gamma_r)}{2} \rightarrow P_i$.
\STATE 3. if $P_i > P_{th}$, $N^* = 2N^*$.
\UNTIL{$P_i \leq P_{th}$}
\STATE \textbf{OUTPUT} $N^*$.
\end{algorithmic}
\end{algorithm}

It can be noted that, when the train location angle $\theta_b = \frac{\pi}{2}$, the adjusting of $N$ would not improve effective beam-forming probability significantly. In that case, $\gamma_l$ or $\gamma_r$ is limited by system structure instead of beam number $N$, which needs further considerations. That is, the optimal beam number $N^*$ searching method can be extended to scenarios with $\theta_b \in (\frac{\pi-\alpha}{2}, \frac{\pi}{2})\cup (\frac{\pi}{2}, \alpha)$.

\section{Application of Location-aware Low-complexity Beam-forming for Massive MIMO System}

\subsection{The Beam Generation Process}
Directional beam can be generated through diverse phase excitations on each element according to \cite{antenna, directional}. Assuming the total generated beam number is $N$ which depends on the BS deployments, and the beam weight of the $i$-th beam ($i=1,2,...,N$) can be expressed as

\begin{equation}\label{weight}
w_i=\sqrt{f_iD_i(\theta_{b})}\tag{12},
\end{equation}
where $f_i=\sum\nolimits_{m=1}^Mf_i(m)$ denotes the power allocation coefficient for the $i$th beam and $f_i(m)$ stands for the actual amplitude excitation on the $m$-th ($m=1,2,...,M$) element for the $i$-th beam. Usually, for a uniform linear array the amplitude excitation is equal on each element. $D_i(\theta_{b})$ is the directivity of the selected $i$-th beam according to the location information $\theta_{b}$.

If we label the phase excitation for $i$-th beam as $\bm{\beta}^m_i$, the corresponding uplink steering vector on each element for an acquired location information $\theta_{b}$ can be denoted as

\begin{equation*}
\bm{e}_i(\theta_{b})=\left [1,e^{j(kd\cdot \cos\theta_{b}+\beta_i^2)}, \ ... \ ,e^{j(M-1)(kd\cdot \cos\theta_{b}+\beta_i^M)} \right ]^T,
\end{equation*}
where $k=\frac{2\pi}{\lambda}$.

\begin{figure}[!t]
\centering
\includegraphics[width=0.48\textwidth]{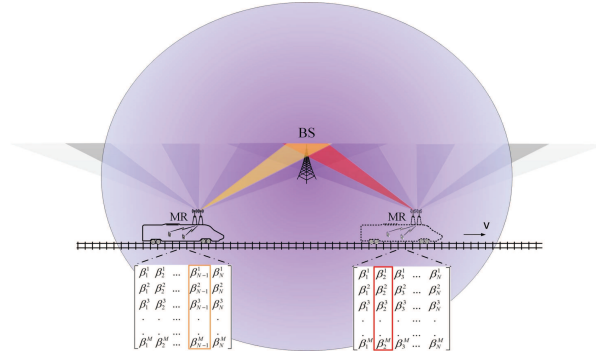}
\caption{The location-aware beam-forming process.} \label{fig:process}
\captionsetup{belowskip=-10pt}
\end{figure}

\begin{rem}
For deployed antenna array, one can off-line calculate and pre-set the phase excitation at the MR according to the different but restricted location information $\theta_{b}$, $\theta_{b}\in \left [\frac{\pi}{2}-\frac{\alpha}{2}, \frac{\pi}{2}+\frac{\alpha}{2}\right ]$. The pre-calculated phase excitation for $i$-th beam is

\begin{equation}\label{phase1}
\bm{\beta}_i=\left [\beta_i^1, \beta_i^2, \ ... \ , \beta_i^M \right ]^T\tag{13},
\end{equation}
and the whole mapper for each beam directed to diverse locations on the rail can be expressed as
\begin{equation}       
\bm{\beta}=\left[                 
  \begin{array}{ccccc}   
    \beta_1^1 & \beta_2^1 & \beta_3^1 & ...  & \beta_N^1\\  
    \beta_1^2 & \beta_2^2 & \beta_3^2 & ...  & \beta_N^2\\  
    \beta_1^3 & \beta_2^3 & \beta_3^3 & ...  & \beta_N^3\\
      \cdot   &   \cdot   &   \cdot   & ...  &   \cdot  \\
    \beta_1^M & \beta_2^M & \beta_3^M & ...  & \beta_N^M\\
  \end{array}\tag{14}
\right].                 
\end{equation}
\end{rem}

\begin{algorithm}[!b]\label{alg:Proc}
\caption{The location-aware beam selection procedure}
\begin{algorithmic}[1]
\label{alg:Proc}
\STATE \textbf{INPUT} $\theta_{b}$.
\STATE {Initialize $\bm{\beta}_c=[\ ]$.}
\REPEAT
\STATE 1. generate N with \textbf{Algorithm 1}.
\STATE 2. acquiring $\theta_{b}$.
\STATE 3. if $\theta_{b}\in \left[ \left(\frac{\pi}{2}-\frac{\alpha}{2}\right)+(i-1)\frac{\alpha}{N}, \left (\frac{\pi}{2}-\frac{\alpha}{2}\right )+i\frac{\alpha}{N}\right ]$, $\bm{\beta}_c=\bm{\beta}_i$.
\UNTIL{$\theta_{b}\ge \frac{\pi}{2}+\frac{\alpha}{2}$}
\STATE {set $\bm{\beta}_c=\bm{\beta}_1$.}
\STATE \textbf{OUTPUT} The phase excitation for current location $\bm{\beta}_c$.
\end{algorithmic}
\end{algorithm}

\subsection{The Beam Selection Process}
The beam selection process is based on the acquired location information. For example, when the train entrances the coverage of one BS for the first time, the right-most beam will be selected to transmit data. The location information will be continuously updated and matched with the phase excitation mapper as the train keeps moving. When the train moves into a new location covered by another beam, the corresponding phase excitation vector will be utilized, which is illustrated as in Fig \ref{fig:process}. Therefore, the whole location-aware beam selection process can be simplified. It looks like as a routing process, where the phase excitation mapper $\bm{\beta}$ is the routing table. The detailed selection process is expressed in Algorithm \ref{alg:Proc}, which requires no CSI detections or ED CCM and therefore, reduces the online computational complexity.


\section{Numerical Results}\label{Sec:NumRes}
In this section, numerical results are presented. Assume that $h=50$m, $f_c = 2.4$GHz, $\lambda = \frac{1}{f_c}$ and $d = \frac{\lambda}{2}$.

\begin{figure}[t]
\centering
\includegraphics[width=0.48\textwidth]{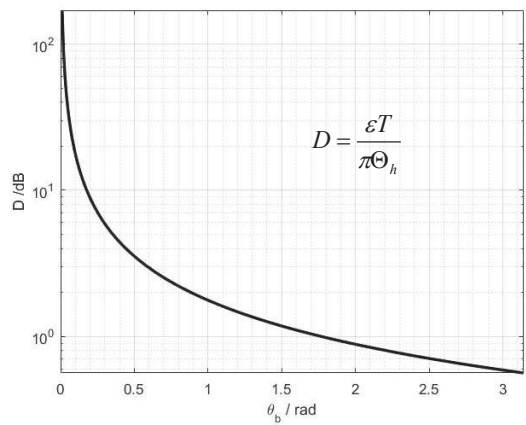}
\caption{The tradeoff between beam directivity and beamwidth.} \label{Fig:Tradeoff}
\end{figure}
\begin{figure}[b]
\centering
\includegraphics[width=0.48\textwidth]{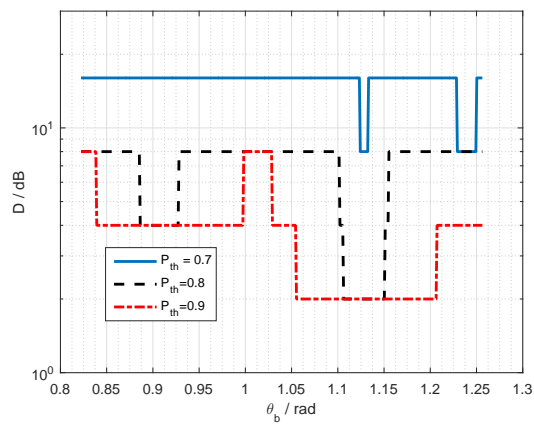}
\caption{The variation of directivity versus base station $\theta_b$ when $P_{th}$ = 0.7, 0.8 and 0.9.} \label{Fig:DversusThetab}
\end{figure}

The tradeoff between $\Theta_h$ and $D$ is shown in Fig. \ref{Fig:Tradeoff}, where for a given beamwidth restricted by the positioning error, the corresponding beam directivity is shown. It can be observed that the beam directivity $D$ decreases with respect to beamwidth $\Theta_h$, where $\Theta_h$ varies from $0$ to $\pi$. It also indicates that the beam directivity increases with total beam number.

As shown in Fig. \ref{Fig:DversusThetab}, when $P_{th}$ = 0.7, 0.8 and 0.9, the variation of directivity versus train location $\theta_b$ has been depicted. It can be observed that the directivity varies with respect to $\theta_b$, but there exists no monotonicity. When the BS is near to the edge of BS coverage, to guarantee the common beam-forming rate, the beam number tends to be relatively large. However, when the BS is near the center of one beam, the beam number tends to be relatively small. It reflects the adaptation of directivity-beamwidth tradeoff in our optimization to guarantee diverse thresholds.

\begin{figure}[t]
\centering
\includegraphics[width=0.48\textwidth]{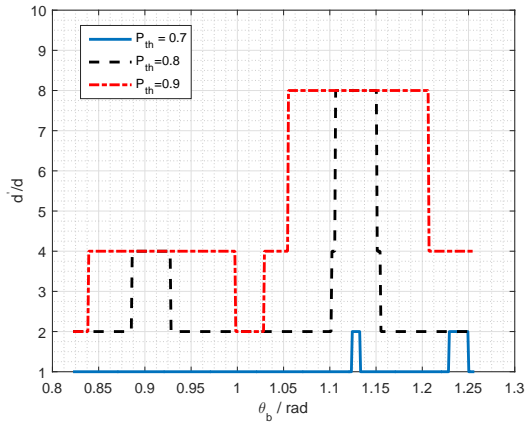}
\caption{The variation of $d$ versus base station $\theta_b$ when the directivity is constant and $P_{th}$ = 0.7, 0.8 and 0.9.} \label{Fig:bVersusThetab}
\end{figure}

\begin{figure}[b]
\centering
\includegraphics[width=0.48\textwidth]{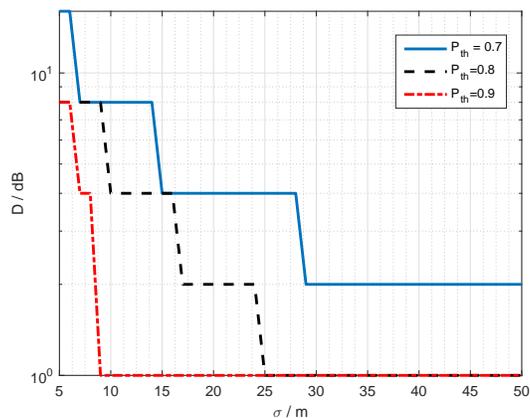}
\caption{The variation of directivity versus positioning error variance $\sigma$ when $P_{th}$ = 0.7, 0.8 and 0.9; $\theta_b=\frac{\pi}{4}$.} \label{Fig:DversusSigma}
\end{figure}

In addition, Fig. \ref{Fig:bVersusThetab} shows that when the directivity is fixed, the variation of $\frac{d}{d^{'}}$ is exactly the inverse of $\frac{N^{'}}{N}$, where $d^{'}$ represents the altered antenna spacing. The numerical results agree with that expressed in $\bold{Remark}$ 1. It also indicates that to guarantee a better robustness, the antenna spacing needs to be designed as large as possible, which reduces the implementation complexity owing to the large space on the top of train.

The variation of directivity versus positioning error variance $\sigma$, when $P_{th}$ = 0.7, 0.8 and 0.9 and $\theta_b=\frac{\pi}{4}$, are shown in Fig. \ref{Fig:DversusSigma}. The directivity decreases linearly with $\sigma$ but increases with $P_{th}$ because larger $\sigma$ or $P_{th}$ means smaller beam number will be adopted so that the switching times among the beams will be decreased. That is, the switching drop probability will be reduced.

\section{Conclusions}

In this paper, we first analyzed the beam-forming design principles for massive MIMO system based on location information and then presented a low-complexity beam-forming implementation scheme in high mobility scenario. Different from conventional beam-forming schemes, our design needs neither acquiring UCCM and DCCM nor ED them, which not only substantially reduces the system complexity and on-line computational complexity, but also possesses a favourable robustness to the CSI estimations. Therefore, the proposed beam-forming scheme can benefit the design of wireless communication system for HST. It is noted that the location information plays a paramount role in our scheme, where the tradeoff between beamwidth and directivity in this scenario and how to maximize direcitivity under diverse positioning accuracies to guarantee efficient transmission are crucial, especially in engineering design of HST wireless communication systems.

\section*{Acknowledgement}

This work was supported by State Key Development Program of Basic Research of China No. 2012CB316100(2) and National Natural Science Foundation of China (NSFC) No. 61321061.
\end{document}